\documentclass[preprintnumbers,article,amsmath,amssymb,floatfix,10pt,prd,onecolumn,
superscriptaddress,nofootinbib]{revtex4-2}
\usepackage{bm}
\usepackage{amsfonts}
\usepackage{latexsym}
\usepackage[latin1]{inputenc}
\usepackage{graphicx}
\usepackage{amsmath}
\usepackage{palatino}
\usepackage{mathpazo}
\usepackage{textcomp}
\usepackage{float}
\usepackage{booktabs}
\usepackage{dcolumn}
\usepackage{ragged2e}
\usepackage{hyperref}
\hypersetup{colorlinks,citecolor=blue}
\hypersetup{colorlinks=true,linkcolor=red,filecolor=magenta,    urlcolor=blue}
\usepackage{amsmath}
\usepackage{xcolor}
\usepackage{orcidlink}
\usepackage{epsfig}
\usepackage[caption=false]{subfig}
\usepackage{commath}

\usepackage{multirow}
\newcommand{\be}{\begin{equation}}
\newcommand{\ee}{\end{equation}}
\newcommand{\bea}{\begin{eqnarray}}
\newcommand{\eea}{\end{eqnarray}}
\newcommand{\eeas}{\end{eqnarray*}}
\newcommand{\beas}{\begin{eqnarray*}}

\newtheorem{remark}{Remark}[section]

\def\jnl@style{\it}
\def\aaref@jnl#1{{\jnl@style#1}}

\def\aaref@jnl#1{{\jnl@style#1}}

\def\aj{\aaref@jnl{AJ}}                   
\def\apj{\aaref@jnl{ApJ}}                 
\def\apjl{\aaref@jnl{ApJ}}                
\def\apjs{\aaref@jnl{ApJS}}               
\def\apss{\aaref@jnl{Ap\&SS}}             
\def\aap{\aaref@jnl{A\&A}}                
\def\aapr{\aaref@jnl{A\&A~Rev.}}          
\def\aaps{\aaref@jnl{A\&AS}}              
\def\mnras{\aaref@jnl{Mon.~Not.~Roy.~Astron.~Soc.}}             
\def\prd{\aaref@jnl{Phys.~Rev.~D}}        
\def\prc{\aaref@jnl{Phys.~Rev.~C}}  
\def\prl{\aaref@jnl{Phys.~Rev.~Lett.}}    
\def\qjras{\aaref@jnl{QJRAS}}             
\def\skytel{\aaref@jnl{S\&T}}             
\def\ssr{\aaref@jnl{Space~Sci.~Rev.}}     
\def\zap{\aaref@jnl{ZAp}}                 
\def\nat{\aaref@jnl{Nature}}              
\def\aplett{\aaref@jnl{Astrophys.~Lett.}} 
\def\apspr{\aaref@jnl{Astrophys.~Space~Phys.~Res.}} 
\def\physrep{\aaref@jnl{Phys.~Rep.}}      
\def\physscr{\aaref@jnl{Phys.~Scr}}       
\def\commat{\aaref@jnl{Comm.~Math.~Phys.}}              
\def\science{\aaref@jnl{Science}}               
\def\cqg{\aaref@jnl{Classical Quant.~Grav.}}            
\def\jpcs{\aaref@jnl{JPCS}}                                     
\def\ijmpd{\aaref@jnl{Int.~J.~Mod.~Phys.~D}}                    
\def\grg{\aaref@jnl{Gen.~Relat.~Gravit.}}               
\def\rpp{\aaref@jnl{Rep.~Prog.~Phys.}}          
\def\npa{\aaref@jnl{Nucl.~Phys.~A}}        
\def\lrr{\aaref@jnl{Living Rev.~Rel.}}                   
\def\jcap{\aaref@jnl{J.~Cosmology Astropart.~Phys.}}    
\def\rmp{\aaref@jnl{Rev.~Mod.~Phys.}}   
\def\epjc{\aaref@jnl{Eur.~Phys.~J.~C}} 
\def\plb{\aaref@jnl{~Phy.~Lett.~B}} 
\def\mpla{\aaref@jnl{Mod.~Phy.~Lett.~A}} 
\def\arxiv{\aaref@jnl{arxiv.org}}

\allowdisplaybreaks[1]

\begin{document}

\color{black}       

\title{Comment on ``Energy conditions in $f(Q)$ gravity"}

\author{Avik De\orcidlink{0000-0001-6475-3085}}
\email{avikde@utar.edu.my}
\affiliation{Department of Mathematical and Actuarial Sciences, Universiti Tunku Abdul Rahman, Jalan Sungai Long,
43000 Cheras, Malaysia}
\author{Loo Tee How\orcidlink{0000-0003-4099-9843}}
\email{looth@um.edu.my}
\affiliation{Institute of Mathematical Sciences, Universiti Malaya, 50603 Kuala Lumpur, Malaysia}

\begin{abstract}
In Phys. Rev. D 102, 024057 (2020), the authors studied energy conditions in $f(Q)$ theory following the same path as researchers handled the energy conditions in the curvature-based modified gravity theories, like $f(R)$ or $f(R,G)$ theories. However, the field equations in the $f(Q)$ theory was not expressed as an effective theory in the literature earlier and in the above mentioned paper \cite{sahoo/ec}, the authors claimed that the pressure and energy density follow a set of energy condition criteria without showing how they came to this conclusion. In this comment, we express $f(Q)$ theory as an effective theory in Friedmann-Lema\^{i}tre-Robertson-Walker (FLRW) universe. And from there, the correct energy conditions are obtained in the traditional (general relativistic) way. Unfortunately, some missing terms in the published work \cite{sahoo/ec} are noticed. 
\end{abstract}
\maketitle

\section{\textbf{Introduction}}

As introduced in the original publication \cite{sahoo/ec}, symmetric teleparallel gravity theory received much attention in the past couple of years. However, unlike its elder brothers, the Levi-Civita connection based general relativity theory and the torsion based metric teleparallel theory, which are already at their maturity, the symmetric teleparallel gravity theory is still at its infant stage. Lots of development and analysis are still going on before we can really estimate its full potential and utilise its strength over its two other counterparts. Some important cosmological applications of this theory can be found in the literature, however, not much works have been done in its geometric aspects. This is one such attempt towards a better understanding of the structure in the background of a spatially flat Friedmann-Lema\^{i}tre-Robertson-Walker (FLRW) spacetime. 

Recently, the covariant formulation of this theory was obtained in terms of the underlying Levi-Civita connection (denoted by a over-circle here) and it was used effectively to study the geodesic deviations, spherically symmetric solutions and the isotropization process \cite{ss, gde, zhao, epjc}
\begin{equation} \label{FE}
f_Q \mathring{G}_{\mu\nu}+\frac{1}{2} g_{\mu\nu} (f-f_QQ) + 2f_{QQ} P^\lambda{}_{\mu\nu} \mathring{\nabla}_\lambda Q = -\kappa T_{\mu\nu}.
\end{equation}
Using this covariant formulation, in the next section we establish the modified $f(Q)$ theory as an effective theory comparable to GR and thus systematically derive the correct energy conditions what was lacking in the earlier literature. In due process, unfortunately, some missing terms in the expressions of the (effective) isotropic pressure and energy density are noticed in the original paper \cite{sahoo/ec} which greatly impact the final results our learned friends obtained. Their methodology and execution are also questionable with the field equations they worked with in the absence of the covariant formulation of the symmetric teleparallel theory at the time of its publication. A more careful handling of this delicate topic is anticipated as carried out in our present article. 


\section{$f(Q)$\textbf{-theory as an effective theory in FLRW universe}}\label{sec2}
As discussed in \cite{ecfr/2018, ec, ecgen}, we express the field equation of any modified gravity theories in the form
\begin{equation} \label{mg}
F(\psi^i)\left(\mathring{G}_{\mu\nu}+H_{\mu\nu}\right)=\kappa T_{\mu\nu},
\end{equation}
where as usual the Einstein tensor is denoted by $\mathring{G}_{\mu\nu}$, the geometrical quantity $H_{\mu\nu}$ contains all the geometrical modifications provided by the considered theory and $\psi^i$ denotes the curvature invariants or gravitational fields contributing to the dynamics. To reawaken GR, we need $F(\psi^i)=1$ and $H_{\mu\nu}=0$. The perfect fluid type energy-momentum tensor is given by the form
\begin{equation} \label{T}
T_{\mu\nu} = (p+\rho)u_{\mu} u_{\nu}+pg_{\mu\nu},\end{equation}
where $p$ denotes the isotropic pressure and $\rho$ the energy density.

As an alternative form of (\ref{mg}), it is a common practice to rewrite the field equations of modified gravity theories as an effective theory by bringing the additional geometrical contributions in the right hand side as follows:
\begin{equation} \label{effective}
\mathring{G}_{\mu\nu}=\frac{\kappa}{F} T^{eff}_{\mu\nu},
\end{equation}
where $ T^{eff}_{\mu\nu}= T_{\mu\nu}-\frac{F}{\kappa}H_{\mu\nu}$. The latter term can be coined as $T^{DE}_{\mu\nu}$, the energy-momentum tensor of the hypothetical fluid produced from geometrical effect, and this quantity is not restricted in terms of curvature effect only as we are going to see in this article for the symmetric teleparallelism.

We consider the homogeneous and isotropic model of the universe described by the spatially flat FLRW metric, the line element in the Cartesian coordinates (coincident gauge) being given by
\begin{equation} \label{flrw}
ds^2 = -dt^2 + a^2(t)\delta_{ij}dx^i dx^j 
\end{equation}
where $a(t)$ is the scale factor of the universe.

We use the following notations: 
\begin{equation}
u_\mu=(\partial_t)_\mu;\quad H(t)=\frac{\dot a}{a}. 
\end{equation} 
The Einstein tensor and Ricci scalar are expressed as 
\begin{align}\label{eqn:einstein}
\mathring G_{\mu\nu}=&-(2\dot H+3H^2)g_{\mu\nu}-2\dot Hu_\mu u_\nu \\
\mathring R= &6\dot H+12 H^2,\nonumber
\end{align}
where the over-dot denotes the derivative over $t$.
Furthermore,  
\begin{equation}\label{eqn:a00}
\mathring\nabla_\mu u_{\nu}=H(g_{\mu\nu}+u_{\mu}u_{\nu}).
\end{equation}
The only non-vanishing Christoffel symbols
\begin{equation*} 
\mathring{\Gamma}^j{}_{0j} = \frac{\dot{a}}{a}  = \mathring{\Gamma}^j{}_{j0}; 
\quad \mathring{\Gamma}^0{}_{jj} = a \dot{a},\quad (j=1,2,3)
\end{equation*}
can alternatively be expressed as
\begin{equation*}
\mathring{\Gamma}^\lambda{}_{\mu\nu} = -
H(-u^\lambda g_{\mu\nu} + u_\mu \delta_\nu^\lambda + u_\nu \delta^\lambda_\mu +u^\lambda u_\mu u_\nu).
\end{equation*} 
Straightforward calculations give
\begin{align*}
Q_{\lambda\mu\nu}=&-2Hu_\lambda (g_{\mu\nu}+u_{\mu}u_{\nu}) \\
Q_\lambda= &-6Hu_\lambda, \quad \tilde{Q}_\lambda=0 \\
P_{\lambda\mu\nu}=&\frac14H(-4u_\lambda g_{\mu\nu}+u_\mu g_{\lambda\mu}+u_\nu g_{\lambda\mu}-2u_\lambda u_\mu u_\nu).
\end{align*}
Using these data, we compute
\begin{align}
Q=&6H^2  \label{eqn:a10}\\
\nabla_\lambda Q=& -\dot Qu_\lambda=-12H\dot H u_\lambda \label{eqn:a20}\\
\nabla_\lambda QP^\lambda{}_{\mu\nu}=&-12H^2\dot H (g_{\mu\nu}+u_{\mu}u_{\nu}). \label{eqn:a30}
\end{align}
Therefore, in the framework of effective theory, we can rewrite the field equations (\ref{FE}) as 
\begin{equation} 
\mathring{G}_{\mu\nu}=\frac{\kappa}{-f_Q} T^{eff}_{\mu\nu}\label{eff-gr}
\end{equation}
where 
\begin{eqnarray*} \label{FE-3}
\kappa T^{eff}_{\mu\nu}&=&
 \kappa T_{\mu\nu}+\left(\frac{1}{2}f-3H^2f_Q \right)g_{\mu\nu}
-24H^2\dot {H} f_{QQ}(g_{\mu\nu}+u_\mu u_\nu )\\
&=&\kappa T_{\mu\nu}+\left(\frac{1}{2}f-3H^2f_Q \right)g_{\mu\nu}
-2\dot{f}_{Q}H(g_{\mu\nu}+u_\mu u_\nu ), \text{ due to } (\ref{eqn:a10}).
\end{eqnarray*}
After a straightforward calculation,  the effective pressure and effective density are given by
\begin{align}
p^{eff}=&p+\frac{f-6H^2f_Q-4\dot {f}_{Q}H}{2\kappa}\label{peff}\\
\rho^{eff}=&\rho+\frac{6H^2f_Q-f}{2\kappa}\label{rhoeff}.
\end{align}
 
\begin{remark}
At this point, it is clear that the authors of \cite{sahoo/ec} unfortunately missed the term $\frac{6H^2f_Q}{2\kappa}$ in both the effective pressure and energy density.
\end{remark}

\begin{remark}As an alternate way, we can obtain the same $p^{eff}$ and $\rho^{eff}$ as (\ref{peff}) and (\ref{rhoeff}), respectively, from the Friedmann type equations of the $f(Q)$-theory, that is, the $00$ and $11$ components of the field equations (\ref{FE})
\begin{equation}\label{a1}
6H^2f_Q=\frac{f}{2}-\kappa\rho
\end{equation}   
\begin{equation}\label{a2}
(2\dot{H}+6H^2)f_Q+2H\dot{f_Q}=\frac{f}{2}+\kappa p
\end{equation}
We can rearrange (\ref{a1}) and (\ref{a2}) into 
\begin{eqnarray}
3H^2&=&\frac{\kappa}{-f_Q}\left[\rho+\frac{6H^2f_Q-f}{2\kappa}\right]\\
-(2\dot{H}+3H^2)&=&\frac{\kappa}{-f_Q}\left[p+\frac{f-6H^2f_Q-48H^2\dot{H}f_{QQ}}{2\kappa}\right].
\end{eqnarray}
In analogy with GR we get the effective pressure and energy density, exactly same as we obtain in (\ref{peff}) and (\ref{rhoeff}). We believe that these are the equations the original paper \cite{sahoo/ec} aimed for but could not manage to achieve.
\end{remark}

\section{\textbf{Conclusion}}
In the present article we have investigated the modified $f(Q)$-gravity theory in a spatially flat FLRW spacetime and systematically expressed its field equations as an effective theory comparable to the Einstein's field equations in general relativity. We have then developed the energy conditions following the traditional approach. In due process, the published result in \cite{sahoo/ec} has been found to be incorrect with missing terms $\frac{3H^2 f_Q}{\kappa}$ in their effective pressure and energy density resulting into wrong set of energy conditions altogether. We have offered the correction in this comment which is vital in the future development of the symmetric teleparallel theory.

\end{document}